# *Ab initio* simulations of oxygen interaction with surfaces and interfaces in uranium mononitride

**D. Bocharov**[1,2*], **D. Gryaznov**[3], **Yu.F. Zhukovskii**[3], **and E.A. Kotomin**[3]

[1]*Faculty of Computing, University of Latvia, 19 Rainis Blvd., Riga, Latvia*
[2]*Faculty of Physics and Mathematics, University of Latvia, 8 Zellu Str., Riga, Latvia*
[3]*Institute of Solid State Physics, University of Latvia, 8 Kengaraga Str., Riga, Latvia*

**Abstract**

The results of DFT supercell calculations of oxygen behavior upon the UN (001) and (110) surfaces as well as at the tilt grain boundary are presented. Oxygen adsorption, migration, incorporation into the surface N vacancies on (001) and (110) surfaces have been modeled using 2D slabs of different thicknesses and supercell sizes. The temperature dependences of the N vacancy formation energies and oxygen incorporation energies are calculated. We demonstrate that O atoms easily penetrate into UN surfaces and grain boundaries containing N vacancies, due to negative incorporation energies and a small energy barrier. The Gibbs free energies of N vacancy formation and O atom incorporation therein at the two densely-packed surfaces and tilt grain boundaries are compared. It has been also shown that the adsorbed oxygen atoms are highly mobile which, combined with easy incorporation into surface N vacancies, explains efficient (but unwanted) oxidation of UN surfaces. The atomistic mechanism of UN oxidation *via* possible formation of oxynitrides is discussed.

1. **Introduction**

Although uranium mononitride (UN) is considered as a possible fuel material for the future Generation IV nuclear reactors, which possesses several advantages as compared to the commonly used uranium oxide fuels [1], presence of oxygen impurities in nitrides and carbides unavoidably leads to unwanted contamination and further degradation of nuclear fuel [2, 3]. A number of experiments were performed so far, aimed at analysis of oxygen effects on UN properties [2-8], including thin films and polycrystalline samples. Rapid oxidation of UN starts at 250$^o$C being accompanied with the N loss [9]. In fact, it is mostly an oxide phase at 500$^o$C which is observed in x-ray studies. Thus, it is important to understand at atomistic level the mechanism of the initial stage of UN oxidation, in general, and the role of interfaces in these processes, in particular, in order to reduce or eliminate unwanted oxidation process.

The experimental studies employing photoelectron spectrometry have also shown that oxygen contact with UN can lead to the chemical transformation of a surface area to either oxynitrides $UO_xN_y$ or formation of $UO_2$ layer, depending on partial pressures of $N_2$ and $O_2$ [8-10]. The observed experimental UPS peak related to O (*2p*) electrons in $UO_xN_y$ is growing with increasing of oxygen content at higher binding energies than the

---

[*] Corresponding author: bocharov@latnet.lv

N (*2p*) peak [10] (even though both are strongly hybridized), in agreement with our calculations [11].

A number of *ab initio* calculations on UN bulk were performed in recent years, mostly within the formalism of Density Functional Theory (DFT) combined with plane wave basis sets [12-16]. Several studies were focused specifically on bulk defects including volatile fission products [17] and O impurities in interstitial and substitutional positions [11, 16, 18]. It was shown, in particular, that O impurities affect visibly UN properties, *e.g.*, mobility of intrinsic defects and related thermal creep [18]. First *ab initio* calculations on the (001) surface and its reactivity have been performed only recently [19-23]. We considered moderate concentrations of N vacancies and O atoms incorporated therein which allowed us to study the early stages of $UO_xN_y$ formation. Our previous calculations on O reactivity upon the (001) surface clearly have shown a possibility of spontaneous breaking of the $O_2$ chemical bond after molecular adsorption [21], strong O adatom chemisorption atop the surface U atoms [20] (typical for metallic surfaces), and energetically favorable incorporation of O adatoms into surface nitrogen vacancies [23]. In these calculations, defect energetics, atomic structures, effective atomic charges, electronic charge densities and density of states were calculated and analyzed.

In the present study, we have extended these models by including the temperature effect, due to variation of the chemical potentials of N and O for the two (001) and (110) surfaces, and also have simulated oxygen behavior at the symmetric tilt grain boundary.

## 2. Methodology

### 2.1. Models and computational parameters

In our large-scale spin-polarized calculations on UN surfaces the VASP 4.6 code has been employed [24, 25]. The projector augmented wave (PAW) [26, 27] method together with the ultra-soft pseudopotentials combined with the non-local exchange-correlation functional (PW91) within generalized gradient approximation [28] have been also used.

We have modeled the UN surfaces using symmetric 2D slabs (Fig. 1) of different thicknesses periodically repeated along the *z* axis and separated by large vacuum gaps (~40 Å). The 2×2 and 3×3 extended supercells correspond to 25% and 11% concentrations of surface defects (adsorbates or vacancies), respectively. The 3D periodic supercell of 15.40×4.87×34.13 Å$^3$ dimensions has been used to model the grain boundary (GB) (Fig. 2) representing a bi-crystal with the total number of 160 atoms. So far, there are no clear experimental evidences on the atomic structure of GBs in UN. In our calculations we have chosen the (310)[001](36.8º) tilt GB observed theoretically and experimentally in other rock-salt materials (see [29] and references therein). We present below the results for the energies of N vacancy formation and O atom incorporation for three different positions at the GB (Fig. 2).

The cut-off energy in plane wave calculations was fixed at 520 eV throughout all the calculations. The Monkhorst-Pack *k*-point mesh of 8×8×1 and 4×4×4 [30] for integration in the Brillouin zone has been used for the slab and GB calculations, respectively. The electronic occupancies were determined following the method of Methfessel and Paxton [31]. The effective charges were calculated using the Bader topological analysis [32, 33]. As stated in experimental study on magnetic properties of UN thin films [34], there is no "bulk-like" antiferro-magnetism in such films since these experiments showed that the magnetic susceptibility of UN thin films decreases with temperature. On the basis of these experimental observations, we have chosen the ferro-magnetic state for all our calculations performed for the self-consistent (relaxed) atomic magnetic moments without spin-orbit effects. Additional information on computational parameters in our calculations can be found elsewhere [20-23].

The defect-free surface energies have been calculated as described in our previous paper [19], the O atom migration energy is calculated as the difference of total energies in equilibrium and transition states of an O atom.

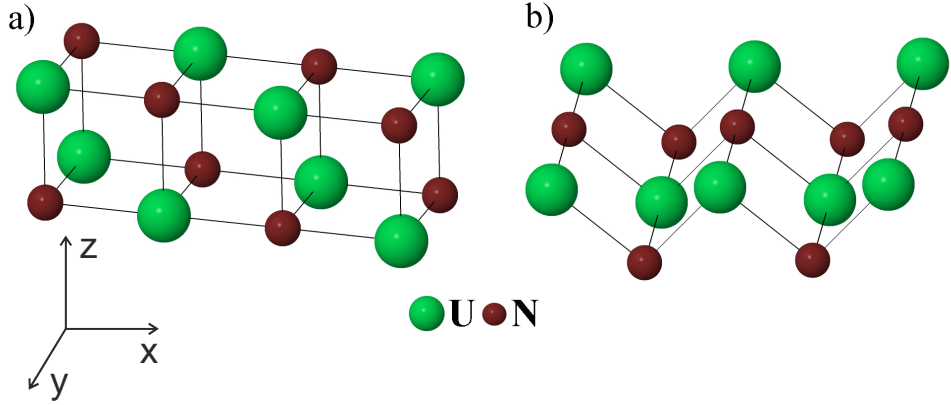

**Fig. 1 (Color online).** The slab models for the UN (001) (a) and (110) (b) surfaces (only the two outermost layers are shown).

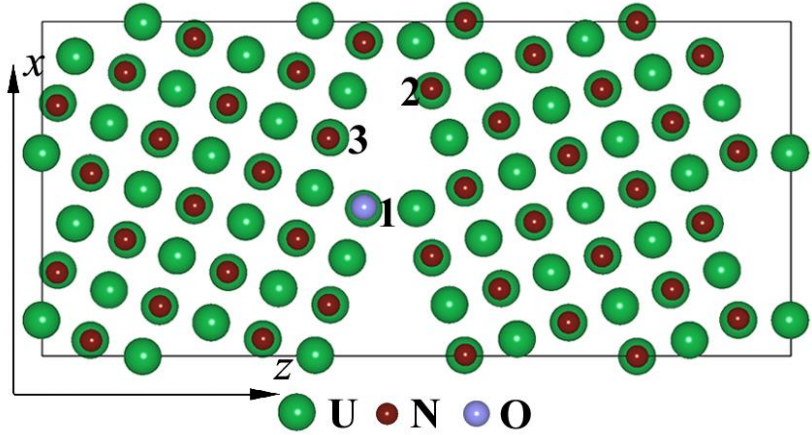

**Fig. 2 (Color online).** The cross-section of the (310)[001](36.8°) tilt GB supercell [29] (15.40 Å × 4.87 Å × 34.13 Å with the oxygen atom incorporated into one of three possible positions (site 1 is shown here, see the text for details).

### 2.2. Defect formation energies

Considering oxygen-rich conditions and the formation of $UO_xN_y$ layers, we suppose that the oxygen chemical potential equals $\frac{1}{2}E_{tot}^{O_2}$, where $E_{tot}^{O_2}$ the total energy of a free $O_2$ molecule. Note that the properties of $O_2$ molecule are poorly reproducible by the DFT [23,35,36]. Thus, we rely on the approach suggested in [37], in order to improve the calculated value of chemical potential of O atom ($\mu_O^0(T)$) important for defect formation energies. In this approach, the total energy of an oxide compound (*e.g.*,

MgO) is used as the reference instead of $O_2$ molecule. Then, the standard Gibbs free *incorporation energy* of O atom into the pre-existing N vacancy in the outermost (surface) layer reads

$$\Delta G_I^O(T) = \frac{1}{2}\left(E_{O\_inc}^{UN} - E_{def}^{UN} - 2\mu_O^0(T)\right), \quad (1)$$

where

$$\mu_O^0(T) = E^{AO} - E^A - \Delta G^{AO}(T^0) + \Delta\mu_o(T). \quad (2)$$

Here $E^{AO}$ and $E^A$ are the total energies of a binary oxide and corresponding metal, $T^0$ the reference (room) temperature, $\Delta G^{AO}(T^0)$ the binary oxide formation energy at standard temperature taken from a thermodynamic database, $\Delta\mu_o(T)$ represents the variation of the chemical potentials for $O_2$ with respect to room temperature [38]. The errors in calculations of defect formation energies in this approach using different binary oxides estimated in [37] are normally much smaller than those expressed the chemical potentials *via* values of $E_{tot}^{O_2}$ calculated using the DFT method. To estimate $\mu_O^0(T)$, we have used MgO and Mg metal as reliable reference materials. Recently, similar approach was successfully used [39] in calculations of formation energies of vacancies in complex perovskite oxides. Then, the standard N chemical potential was determined from the relation $\mu_N^0 = \mu_{NO}^0 - \mu_O^0$, where the chemical potentials of both nitric oxide and oxygen are temperature dependent and require use of thermodynamic data. The standard Gibbs free *formation energy* of N vacancy in the outermost (surface) layer reads

$$\Delta G_F^N(T) = \frac{1}{2}\left(E_{vac}^N - E^{UN} + 2\mu_N^0(T)\right). \quad (3)$$

Here $E_{vac}^N$ is the total energy of UN surface with one nitrogen vacancy, $E^N$ the total energy of defect-free UN surface. Note that the prefactors ½ and 2 in both Eqs. 1 and 3 arise simply due to use of symmetric slabs with defects on its both sides. We suppose in the present study the variation of defect formation (incorporation) energy on temperature, due to the chemical potential of nitrogen (oxygen) only, ignoring the phonon contribution in the solid phase.

## 3. Main results

### 3.1. Defect-free (001) and (110) surfaces

According to Tasker's analysis [40], the (001) surface is expected to have the lowest surface energy for the rock-salt (*fcc*) compounds like MgO and UN. However, one could expect facets with different crystallographic orientations in nano-particles and polycrystalline materials. Therefore, additional calculations are required for other surfaces, in order to check the validity of our results. In the present paper, we consider the (110) surface in addition to earlier studied (001) surface (Fig. 1). The (110) surface is characterized by a reduced coordination number of surface atoms (4 *vs.* 5 at the (001) surface) and smaller interlayer distances along the *z* axis. Thus, the calculated values of surface energies, atomic O binding energies, the Gibbs free formation energies of N-vacancies and O atom incorporation energies, are discussed below for the (110) surface and compared to those for the (001) surface.

The surface energy $E_{surf}$ is given in Table 1 as a function of the number of layers in the slabs for both defect-free surfaces. The lattice relaxation energies turned out to be quite small, ~0.03 eV. Depending on the slab thickness, $E_{surf}$ is by ~0.5-0.7 J·m$^{-2}$ larger for the (110) surface as compared to the (001) one. Besides, $E_{surf}$ for the (110) surface decreases by 0.15 J/m$^2$ with the slab thickness (from 1.98 to 1.83 J/m$^2$) compared with 0.22 J/m$^2$ (from 1.44 to 1.22 J/m$^2$) for the (001) surface. This confirms the energetic preference of the (001) surface, at least at 0 K.

Table 1. Surface energies $E_{surf}$ (in J·m$^{-2}$) as well as effective atomic charges (in $e^-$) on surface N ($q_N^{eff}$) and U ($q_U^{eff}$) atoms on the defect-free UN (001) and (110) surfaces and UN bulk (U,N charges ±1.69 e).

| Number of layers in a slab | (001) | | | (110) | | |
|---|---|---|---|---|---|---|
| | $E_{surf,}$ | $q_N^{eff}$ | $q_U^{eff}$ | $E_{surf,}$ | $q_N^{eff}$ | $q_U^{eff}$ |
| 5 | 1.44 | -1.65 | 1.68 | 1.98 | -1.55 | 1.46 |
| 7 | 1.37 | -1.67 | 1.74 | 1.93 | -1.55 | 1.48 |
| 9 | 1.29 | -1.67 | 1.68 | 1.88 | -1.55 | 1.49 |
| 11 | 1.22 | -1.68 | 1.72 | 1.83 | -1.55 | 1.48 |

The effective Bader atomic charges $q_{U(N)}^{eff}$ for the two surfaces demonstrate stronger U-N bond covalence at the (110) surface as compared to the bulk UN. This is also in contrast to the (001) surface, where the effective charges are close to the bulk values. The differences in effective charges for the two surfaces could be explained by different reconstruction mechanisms of the surfaces. On the (110) surface, the outermost U and N atoms are displaced inwards the slab center, by ~0.042 and 0.014 Å, respectively. In contrast, on the (001) surface the U atoms move inwards the slab center (~0.046 Å), while N atoms are shifted (~0.024 Å) in the opposite direction. The structural properties are also the reasons of the differences in binding energies of O atoms on the two surfaces discussed below.

### 3.2. Oxygen adsorption and migration upon the defect-free surfaces

The binding energies $\Delta E_{bind}^O$ of O adatoms (Table 2) atop surface U and N atoms read

$$\Delta E_{bind}^O = \frac{1}{2}\left(E^{UN} - E_{O\_inc}^{UN} + 2\mu_o(0)\right). \tag{4}$$

The standard chemical potential of oxygen $\mu_O^0$ has been calculated at 0 K according to Eq. 2. For both the surfaces, the values of $\Delta E_{bind}^O$ are larger if O atom adsorbs atop the U atoms than atop the N atoms (by almost 1.9 eV for the (001) and 2.2 eV for the (110) surface). Also, as the supercell size is increased from 2×2 to 3×3 (smaller O coverage), $\Delta E_{bind}^O$ is slightly increased. The values of $\Delta E_{bind}^O$ are larger (by 0.1-0.4 eV) for the (110) surface as compared to the (001) one. The (110) surface reconstruction mechanism results in reduced lateral interactions between the adsorbed O atoms. On the other hand, the adatom effective charges are practically the same for both the (001) and (110) surfaces (Table 2).

**Table 2**. The binding energies $\Delta E_{bind}$ (in eV) and effective charges $q_O^{eff}$ (in $e^-$) for O atom adsorption atop U or N surface atoms on 7 layer UN(001) and (110) slabs.

| Surface and supercell size | | U | | N | |
|---|---|---|---|---|---|
| | | $\Delta E_{bind}$ | $q_O^{eff}$ | $\Delta E_{bind}$ | $q_O^{eff}$ |
| (001) | 2×2 | 4.80 | -1.09 | 2.88 | -1.17 |
| | 3×3 | 4.87 | -1.09 | 2.95 | -1.18 |
| (110) | 2×2 | 5.19 | -1.09 | 3.02 | -1.18 |
| | 3×3 | 5.21 | -1.10 | 3.28 | -1.18 |

Three main *migration paths* of adsorbed O atom upon the UN (001) surface have been identified (Fig. 3 and Table 3): (1) hops between the U atoms along the *x*- or *y*-axes over the N atoms, (2) between the two neighboring U atoms along the (110) direction, or (3) between neighboring N atoms along the (110) direction. The migration energy is calculated as the difference between $\Delta E_{bind}$ for O atom adsorbed above a surface atom and that at the saddle point (with a full lattice relaxation).

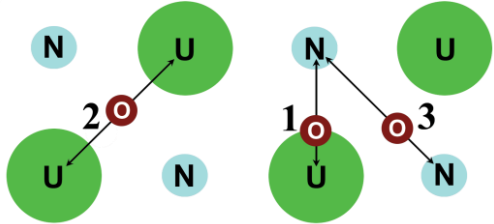

**Fig. 3 (Color online)** Three different oxygen migration paths upon the UN(001) surface (top view).

The results of adsorbed O migration energies calculated for different slab thicknesses and supercell sizes are summarized in Table 3. $\Delta E_{bind}$ energies were calculated at five points along the O migration path in the 2×2 supercell and only two points in the 3×3 supercell (to reduce computational costs). In both cases, the most favorable migration path appeared to be between the nearest surface U atoms (migration path 2). The corresponding migration barrier (0.35 eV for the 5-layer slab and 0.26 eV for the 7-layer slab) indicate high mobility of adsorbed O atoms along the UN (001) surface. The migration barriers along the other two migration paths are much larger (1.93-2.05 eV and 1.31-1.69 eV for migration paths 1 and 3 shown in Fig. 3).

Unlike the (001) surface with alternating N and U atoms along the *x*- and *y*-axes, the (110) surface consists of alternating *rows* of N and U atoms (Fig. 1). It means that the migration of O atoms upon the (110) surface between nearest U atoms (this path is lowest in energy, as discussed above) can occur either in one dimension (inefficient for encountering surface N vacancies) or through periodic O jumps to sub-surface U atoms.

As a result, the oxygen atom on the (110) surface will be less mobile than on the (001) one. Thus, our further consideration of migration processes is focused on the (001) surface only.

**Table 3**. Oxygen migration energies for the three migration paths (Fig. 2).

| Supercell size: | 2×2 | | 3×3 | |
| --- | --- | --- | --- | --- |
| Number of atomic layers: | 5 | 7 | 5 | 7 |
| Path 1 | | | | |
| Migration barrier | 2.05 | 1.92 | 2.02 | 1.92 |
| Path 2 | | | | |
| Migration barrier | 0.35 | 0.26 | 0.39 | 0.36 |
| Path 3 | | | | |
| Migration barrier | 1.69 | 1.67 | 1.63 | 1.56 |

### 3.3. Oxygen incorporation into the N vacancy

The surface U vacancy reveals much smaller formation energy than the N vacancy [22]. This is almost independent of the slab thickness, but decreases slightly, by ~0.06 eV, when increasing the supercell size from 2×2 to 3×3. Using Eqs. (1-3), the standard *Gibbs free formation energy* is plotted in Fig. 4 for the two UN surfaces and GB as a function of temperature. (Different positions of the N vacancy at the GB show, however, very close formation and incorporation energies, thus, we consider here only position 2). (Note that the value of $\Delta G_F^N$ would decrease by 0.5 eV in a pure N$_2$ atmosphere surrounding the surface.) The difference in $\Delta G_F^N$ between the (001) and (110) surfaces is considerable, 0.7 eV (fig. 4) and, most importantly, $\Delta G_F^N$ decreases by 0.4 eV as the temperature increases from 400 to 700 K. The formation energy at the GB lies in-between that for the two surfaces.

The interaction of O atom with the surface could be characterized by the *incorporation energy* $\Delta G_I^O$ (the energy gain due to an O atom occupation of a pre-existing vacancy as given by Eq. 1). Its negative value means that the reaction is exothermic and thermodynamically favorable. Significantly negative values were observed earlier for the (001) surface at 0 K [23], this is an indication of the potential high efficiency of UN oxidation process.

Fig.4 demonstrates that the Gibbs free incorporation energies for the two surfaces differ by ~0.4 eV, these are only slightly dependent on the slab thickness (incorporation into the vacancies on the (001) surface is energetically more favorable). It *increases* with temperature by ~0.4 eV in the temperature range from 400 to 700 K but remains still negative (process energetically favorable). The oxygen incorporation energy for the GB is smaller than for both surfaces, but remains still negative, even at 700 K.

Another important parameter is the *solution energy* $\Delta G_F^N + \Delta G_I^O$, taking into account the vacancy formation energy [23]. For both surfaces and the GB it remains negative as well. We, thus, conclude that O atom penetration into the UN is thermodynamically favorable process which confirms an important role of UO$_x$N$_y$ layers in the UN oxidation process.

Lastly, the adsorbed O atom has to overcome the energy barrier while penetrating into the nearby surface vacancy. We have simulated a series of possible O drop-in trajectories, estimated this energy barrier and analyzed the corresponding electronic charge redistribution. In Fig. 5, the diference electron charge distribution is plotted for O adsorbed atop U atom with neighboring N vacancy. The electronic density tail parallel to the surface between the O atom and N vacancy indicates at a certain energy barrier for incorporation of negatively charged O atom into the N vacancy. (The seven layer slab chosen for these configurations is sufficiently thick to exclude the spurious interaction between the adsorbed O atoms across the slab. On other hand, the interaction between the O atoms is stronger for the 2×2 supercell (Fig. 5a). Simulating different possible trajectories of adsorbed O atom towards the surface vacancy, we found the energy barrier of 0.5 eV (comparable with oxygen migration energy).

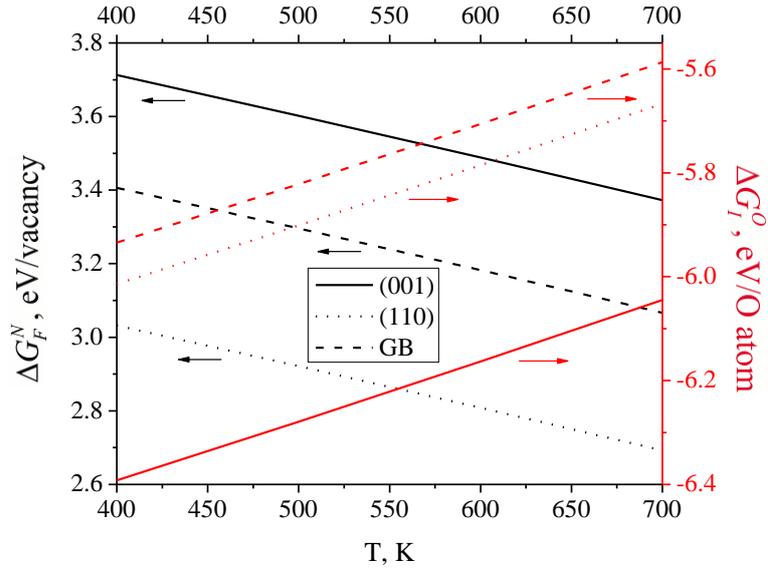

**Fig. 4. (Color online)**. The standard Gibbs free formation energy of N vacancy (black curves) and the incorporation energy of O atom into the surface N vacancy (red curves) as a function of temperature for the (001), (110) surfaces and GB (position 2 in Fig. 2). The supercell size and slab thickness are 3×3 and 7 planes, respectively.

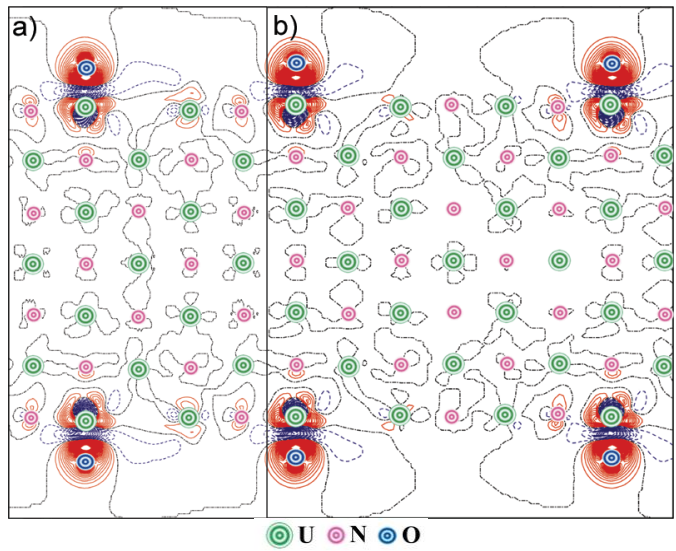

**Fig. 5**. **(Color online)**. The 2D sections of the electron charge density re-distributions $\Delta\rho(\mathbf{r})$ for O atoms adsorbed atop U atom for 2×2 (a) and 3×3 (b) supercells upon the seven-layer UN(001) slab. The N vacancy is at the shortest distance to U atom. Function $\Delta\rho(\mathbf{r})$ is defined as the total electron density of the interface containing adsorbed O atom minus the densities of substrate and adsorbate with optimized interfacial geometry. Solid (red) and dashed (blue) isolines correspond

to positive and negative electron densities, respectively. Dot-dashed black isolines correspond to the zero-level.

## 4. Conclusions

Using the DFT approach, properties of the defect-free and defective UN (001) and (110) surfaces as well as (310)[001](36.8°) tilt grain boundaries have been compared. For the (001) surface, the N vacancy formation energy has been found to be larger than that for the (110) surface, however, the oxygen incorporation energy into the N vacancy at the (001) surface is energetically more favorable. The N vacancy formation energy at the grain boundaries lies in-between those for the two surfaces whereas the O incorporation energy at the vacancy in GB is smaller than at both surfaces (but remains still energetically favorable). Lastly, the O solution energies (incorporating also N vacancy formation cost) in all three cases are predicted to be very close and negative (-2.5 eV), thus the oxidation process is energetically favorable at all studied temperatures.

Detailed study of adsorbed oxygen atom migration upon the UN (001) surface and its penetration into the surface N vacancy has been performed. We have demonstrated quite high mobility of O atoms along this surface, with low migration barriers and a relatively small additional energy barrier for O penetration into the N vacancy. Based on the present and our previous studies [19-23], we have estimated energetics of incorporation mechanism at initial stages with formation of oxynitrides including: (*i*) chemisorption of $O_2$ molecule, (*ii*) spontaneous dissociation of adsorbed $O_2$ molecule, (*iii*) favorable adsorption of released oxygen adatoms atop the U atoms; (*iv*) high mobility of O atoms along the surface, (*v*) low-barrier incorporation of oxygen adatoms from the position atop U atoms into the N vacancies, *(vi)* stabilization of O atoms in surface N vacancies.

**Acknowledgements**

This work is partly supported by the European Commission FP7 project *F*-BRIDGE and ESF project No. 2009/0216/1DP/1.1.1.2.0/09/APIA/VIAA/044. The authors sincerely thank A. L. Shluger and K. McKenna for help with grain boundary model and R. A. Evarestov, R. J. M. Konings, Yu. A. Mastrikov, S. Piskunov, P. Van Uffelen for valuable suggestions and many stimulating discussions.